# Identical chaotic synchronization and bidirectional message transmission in incoherently coupled semiconductor laser diodes

I. V. Koryukin

*Institute of Applied Physics RAS, 46 Ulyanov Street, 603950 Nizhny Novgorod, Russia*

## Abstract

A chaos-based communication scheme allowing simultaneous bidirectional message transmission (Opt. Lett. **32**, 403, 2007) is investigated numerically. Incoherent feedback and coupling case is analyzed, which is expected in real long-distance optical communication systems. It is shown that identical synchronization of chaotic laser waveforms and bidirectional message transmission are possible as in the coherent coupling case. However, the chaotic regime at incoherent feedback and coupling is quite different. It is regular destabilized relaxation oscillations with the chaotic envelope. Such dynamics leads to restriction of the transmitting signal bit rate by a portion of relaxation oscillations frequency.

## I. INTRODUCTION

Chaotic synchronization of nonlinear oscillators has been intensively studied in the recent years as a base for secure communication systems [1,2]. Special interest has been focused on the synchronization of semiconductor lasers which are key elements of available optical communication technology and potential devices for chaos-based optical systems [4-7]. In semiconductor lasers fast high-dimensional chaotic dynamics is easily accessible under different external perturbations including optical or optoelectronic feedback. Such a broadband chaotic carrier allows encoding and transmission of information with bit rates in the gigabit per second range both in laboratory experiments [8] and through commercial fiber optic network [9]. Most of the successful experiments have been performed with unidirectional coupling schemes leading to unidirectional information transmission. For bidirectional communication these schemes require two receivers and two transmitters. Moreover, the level of security provided by unidirectional systems can be insufficient, at least in the case of synchronization of single-mode semiconductor lasers [10].

Chaotic synchronization of bidirectionally coupled semiconductor lasers was demonstrated for the first time in Ref. [11]. It was found that, even for almost identical lasers, direct face-to-face coupling leads to spontaneous symmetry breaking and leader-laggard type of synchronization with relative time lag corresponding to the light propagation time between lasers. Like in the case unidirectional coupling, transmission of encrypted message is possible only in one direction through leader to laggard laser. The problem of bidirectional information exchange in face-to-face coupling scheme was overcome just recently [12]. It was proposed to put a partially transparent mirror to the optical pathway connecting the bidirectionally coupled lasers (Fig. 1). For each laser this mirror provides self-feedback as well as light injection from the other laser. Such a simple scheme allows identical synchronization of chaotic laser intensities, without time lag between them given by transmission of the light from one laser to another.

However, only coherent feedback and coupling corresponding to a short distance between the lasers is investigated in Ref. [12]. On the other hand, the linewidth of telecommunication-grade semiconductor lasers is typically greater than a few hundreds of kilohertz, which makes their coherence length too short (less than 1 km) in comparison with usual fiber-optic links. If the proposed scheme is used in real long-distance optical communication systems, both distances between the laser and the mirror are expected to exceed the coherence length of the corresponding laser. Therefore in this paper I analyze the incoherent feedback and coupling case.

## II. MODEL

A model of single-longitudinal-mode semiconductor laser with incoherent delayed optical feedback [13,14] is adapted for the scheme under consideration (Fig. 1) and takes the following dimensionless form:

$$\frac{dI_1(t)}{dt} = \left[N_1 - 1 - k_{f_1} I_1(t - 2\tau_1) - k_{21} I_2(t - \tau_1 - \tau_2)\right] I_1(t),$$

$$\frac{dI_2(t)}{dt} = \left[N_2 - 1 - k_{f_2} I_2(t - 2\tau_2) - k_{12} I_1(t - \tau_1 - \tau_2)\right] I_2(t) \quad (1)$$

$$T \frac{dN_{1,2}(t)}{dt} = P_{1,2} + 1 - N_{1,2}(t) - N_{1,2}(t) I_{1,2}(t).$$

In these equations the indices 1,2 label the first and second laser variables and parameters, $I$ is the laser field intensity, $N$ is the excess free-carrier density, $k_f$ is the feedback strength, and $k_{21}$ ($k_{12}$) is the injection strength for the first (second) laser. The excess pump current $P$ is proportional to $J/J_{th} - 1$, where $J$ and $J_{th}$ are the injection current and its value at the solitary laser threshold, respectively. The reduced time $t$ is measured in units of photon lifetime $\tau_{ph}$, $T = \tau_s / \tau_{ph}$, where $\tau_s$ is the carrier lifetime, $\tau_1, (\tau_2)$ is the transmission time of light from the laser 1 (laser 2) to the mirror.

For numerical simulation of the model equations (1) internal laser parameters are assumed to be identical for both the lasers. They are chosen close to parameters of ref. [12]: $T = 10^3$ ($\tau_s = 1\ ns,\ \tau_{ph} = 1\ ps$), $P = 1.2$ ($J = 2.2 J_{th}$). The coupling and feedback strengths are symmetric in pairs: $k_{f_1} = k_{f_2} \equiv k_f$, $k_{12} = k_{21} \equiv k_c$, while the delay times $\tau_1$ and $\tau_2$ may be different and vary from a few nanoseconds to a few microseconds.

## III. DYNAMICAL REGIMES AT INCOHERENT FEEDBACK AND CHAOTIC SYNCHRONIZATON

Dynamics of a single laser with incoherent feedback has been studied by keeping coupling strength equal to zero $k_c = 0$. Under this condition, both the lasers operate independently. For definiteness behavior of the first laser will be considered. Without feedback the model equations have a single steady state solution $I_1 = P_1$, $N_1 = 1$, corresponding to stable lasing with constant intensity. Small perturbations decay to this steady state via damped relaxation oscillations with frequency $\Omega_R = \sqrt{P/T}$ and damping rate $\theta_R = -(P+1)/T$. For our parameters, relaxation oscillations frequency is about 5.5 GHz. If the feedback strength $k_{f_1}$ increases some critical value, regular small-amplitude oscillations appear around the destabilized steady-state. The frequency of these oscillations is close to the relaxation oscillations frequency. Further increase in $k_{f_1}$ leads to the complication of the intensity behavior and, finally, to the chaotic regime. Typical chaotic behavior of laser intensity is presented in Fig.2. It is clearly seen that this regime is quite different from the coherent collapse regime realized at coherent feedback in the same region of parameters [12]. In the coherent collapse regime, chaotic behavior is the so-called destabilized relaxation oscillations which have irregularities on the time scale of the relaxation oscillations period. On the contrary, in the case of incoherent feedback, undamped relaxation oscillations are quite regular (Fig. 2(c)), but their amplitude perform complex nonperiodic oscillations with much lower frequencies (Fig. 2(a,b)). The main modulation frequency is coincides with the inverse feedback delay time $f_m = 1/2\tau_1 = 10 MHz$ (Fig. 2(a)). Modulation spectrum contains also a number of harmonics of the main modulation frequency (Fig. 2(b)).

Dynamics of a single class B laser with incoherent long-distance feedback was earlier studied in Ref. [14] using the same model. Regular low frequency behavior of the relaxation oscillations envelop was found for the parameters typical for a solid-state laser ($T=10^5$). The period of these low-frequency pulsations was much larger than the feedback delay. In our case of a semiconductor laser, when the parameter $T$ is about 100 times smaller ($T=10^3$), only chaotic behavior is exist. Neither regular nor chaotic pulsations with a period longer that the delay time were found.

Identical synchronization of chaotic laser intensities arises in a threshold fashion when coupling strength exceeds its critical value. The threshold value of $k_c$ is found to be one order of magnitude smaller than the corresponding self-feedback value, thus for the parameters of Fig. 2, $k_c^{th}$ is (1.5…3) $10^{-4}$, when $k_f$ changes from 2.6 $10^{-3}$ to 2.8 $10^{-3}$. Above this threshold, the synchronization regime is qualitatively independent of the ratio of $k_c$ and $k_f$, if the total injection strength $k_f + k_c$ remains constant. For the mirror placed in the central position between lasers ($\tau_1 = \tau_2 \equiv \tau$), it was found that identical synchronization exists in a wide range of delay time from $\tau = 2ns$ up to $\tau = 2\mu s$ (Fig. 3(a)). If the mirror is moved from the centre to an asymmetric position, even very close to one of the lasers, the synchronization is maintained also (Fig. 3(b)). Only the chaotic behavior itself changes slightly: relaxation oscillations envelope is enriched by a new frequency which is inverse to the difference between delay times $\tau_1$ and $\tau_2$. Therefore, without loss of generality, in the study of message transmission that is presented below I shall use symmetric situation with $\tau_1 = \tau_2$ not to worry about the delay time difference compensation.

### IV. BIDIRECTIONAL MESSAGE TRANSMISSION

A simplest way for the message transition in coupled semiconductor lasers is modulation of the pump current. To check response of the system to such modulation the rectangular pulse was added to the pump current of the SL1. Pulse amplitude is assumed to be small compared to the pump current itself and equal to 0.05$J$, pulse width is 2ns. Temporal desynchronization of the lasers under this pulse action is shown in Fig. 4. Both the relaxation oscillations frequency and the envelope of the oscillations are different for two lasers (Fig. 4(a)). Synchronization error ($I_1(t) - I_2(t)$) differs from zero not only under the pulse action, but damping slowly during an appreciable time after the pulse (Fig. 4(b)). Damping time is about several nanoseconds and is almost independent of the pulse amplitude and width as well as on the majority of laser parameters ($P, \tau_1, \tau_2, k_f, k_c$).

This behavior is quite different from the behavior of synchronization error at the coherent feedback and coupling, where desynchronization was observed only during the application of the pulse to the laser [12]. Obviously, such a difference is a consequence of very different dynamical regimes in the coherent and incoherent cases.

It seems that lack of coincidence between the modulation pulse and the synchronization error, long decay of this error after the pulse, will hamper signal extraction and restrict sufficiently maximal bit rate of the transmitted signal. However, it was found that application to the synchronization error signal a low-frequency digital filter with the cutoff frequency smaller than the relaxation oscillations frequency may solve this problem. In Fig. 4(c) filtered synchronization error is presented together with initial modulation pulse. It is surprising, but duration of the recovered pulse is close to the duration of the initial one and practically unaffected by long synchronization error tail.

Maximal reachable bit rate of the transmitted signal has been evaluated. Pseudorandom bit sequence was applied to the first laser pump current and restored from the synchronization error signal. Different amplitudes of the modulation signal $\Delta J$ were used in the range between $0.01J$ and $0.05J$. System parameters were varied around their typical values ($T = 10^3$, $P = 1.2$, $\tau_1 = \tau_2 = 50$ ns, $k_f = k_c = 1.5 \times 10^{-3}$). It was found that for these parameters and the modulation amplitude $0.05J$ unerring recovering of the initial bit sequence of 1GBit/s is possible. High bit rates are accessible at low amplitudes of the modulation. However, this way of bit rate increasing is restricted by the noise level of the lasers themselves as well as by the channel noise. An increase in $P$ or $k_f$ also leads to rise in the bit rate of the signal transmitted without errors. For instance, at $P = 2$ maximal bit rate is 2.5Gbit/s ($\Delta J = 0.05J$). Unlike the above mentioned parameters, increase in the parameter $T$ results in low accessible bit rates. Detailed investigation has shown that the critical role is played by relaxation oscillations frequency. It was found that minimal duration of the modulation pulse should be 2...3 times greater that the period of relaxation oscillations. Therefore, all variations of parameters leading to increasing the relaxation oscillations frequency also increase the maximal bit rate of the transmitted signal.

Simultaneous modulation of both the laser pump currents ($\Delta J_1$ and $\Delta J_2$) by applying two independent pseudorandom bit sequences allows bidirectional message transmission. An example of the difference between the original messages (modulation signals) $\Delta J_1(t) - \Delta J_2(t)$ and the corresponding signal, subtracted from the synchronization error is presented in Fig. 5 for the messages bit rate of 1Gbit/s. It is clearly seen that subtracted signal correctly reproduces the original one. So, the chaos-based communication scheme proposed in ref. [12] for the coherent case operates also at incoherent feedback and coupling. It can be used for interchange of encrypted key through a public channel with the length exceeding the coherence length of coupling lasers.

All the above unidirectional results concerning the maximal reachable bit rate and its dependence on the parameters are true for the bidirectional transmission case. Therefore, transmission bit rates of several Gbit per second are quite accessible.

## V. CONCLUSIONS

It is shown that in the considered scheme with incoherent feedback and coupling identical synchronization of chaotic laser waveforms exists and simultaneous bidirectional message transition is possible like in the coherent coupling case. However, the chaotic regime at incoherent feedback/coupling is quite different. It is regular destabilized relaxation oscillations with the chaotic envelope. Such dynamics leads to restriction of the transmitting signal bit rate by a portion of relaxation oscillations frequency. Nevertheless, this restriction is inessential for the proposed use of this scheme for exchange of an encrypted key through a public channel.

### ACKNOWLEDGMENTS

This research was supported by the Russian President Program for Support of Leading Scientific Schools (Grant No. 1931.2008.2).

**Figure Captions**

Fig. 1. Schematic setup for bidirectional coupling of two semiconductor lasers (SL1, SL2) through a partially transparent mirror M; OF: optical fibre.

Fig. 2. Temporal traces of the intensity of single uncoupled ($k_c = 0$) laser with incoherent optical feedback (SL1). Three different time scales are shown, the fastest one corresponding to relaxation oscillations (c), the slowest oscillations have the period of delay (a), time is in nanoseconds. The parameters are $T = 10^3$, $P = 1.2$, $\tau_1 = 50$ ns, $k_{f1} = 3 \times 10^{-3}$.

Fig. 3. (Color online) Identical synchronization for the mirror at the central position: $\tau_1 = \tau_2 = 50$ ns (a), and near one of the lasers: $\tau_1 = 91$ ns, $\tau_2 = 9$ ns (b). SL1 output is the low trace (black), SL2 output is the upper trace (grey, red online) and shifted vertically for the convenience. In (b) time lag between lasers is compensated. $k_f = 1.5 \times 10^{-3}$, $k_c = 1.5 \times 10^{-3}$, other parameters are as in Fig. 2.

Fig. 4. (Color online) Desynchronization due to pump current modulation; SL1 (black) and SL2 (grey, red online) output (a), synchronization error (b), modulation pulse (black) and filtered synchronization error (grey, red online) (c). The parameters are as in Fig. 3a.

Fig. 5. Message decryption at bidirectional transmission; initial message (a) and message subtracted from the synchronization error (b). The parameters are $\Delta J = 0.05 J$, $T = 10^3$, $P = 1.2$, $\tau_1 = \tau_2 = 50$ ns, $k_f = 1.5 \times 10^{-3}$, $k_c = 1.5 \times 10^{-3}$.

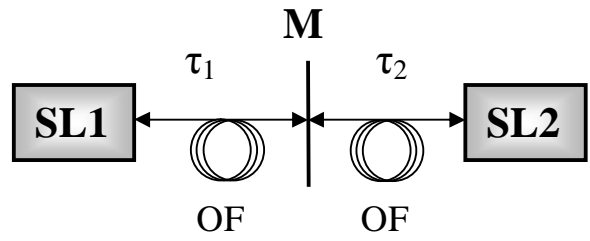

Fig. 1.

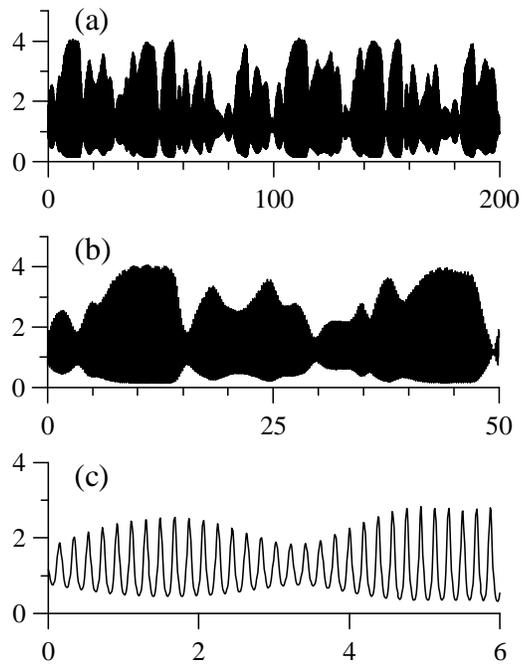

Fig. 2

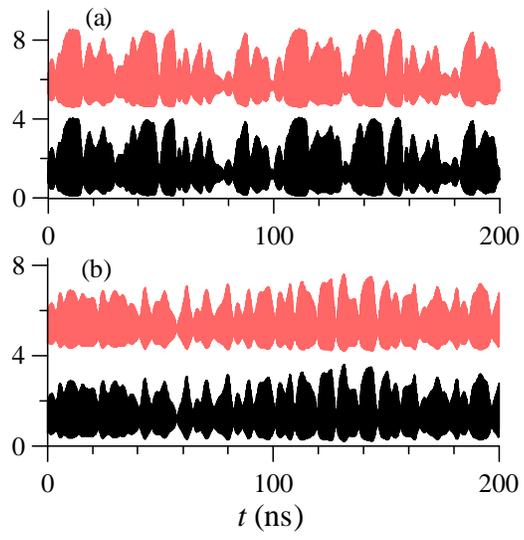

Fig. 3

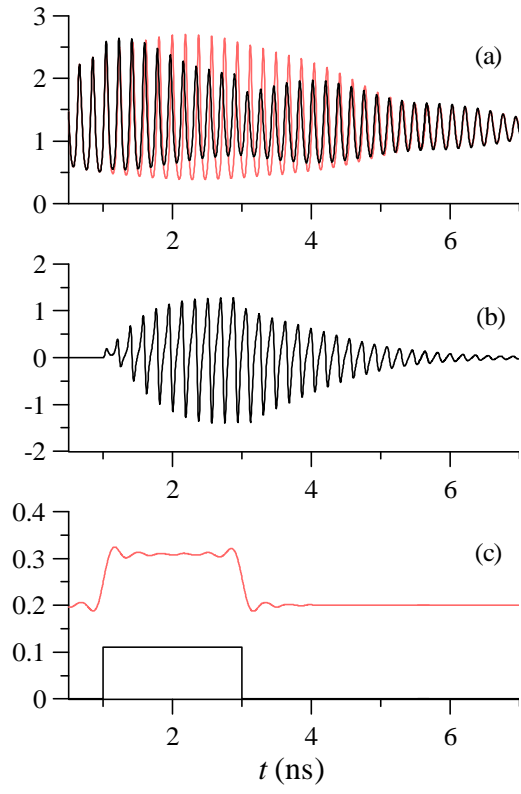

Fig. 4

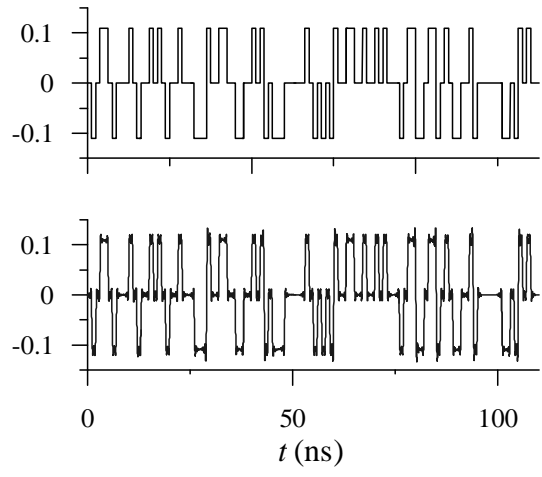

Fig. 5